\documentclass[aip,reprint]{revtex4-1}

\draft 

\usepackage{newtxtext,newtxmath}
\usepackage{color}
\usepackage{amsmath}
\usepackage{graphicx,float}  
\usepackage{epsfig, subfigure}
\usepackage{mathrsfs}
\usepackage{bm}        
\usepackage{amsfonts}
\usepackage{amssymb}
\usepackage{mathrsfs}
\usepackage{natbib}

\usepackage{epsfig,subfigure}
\usepackage{multirow}
\usepackage{caption}
\usepackage{verbatim}  
\begin{document}



\title{Similarity model for corner roll in turbulent Rayleigh-B{\'e}nard convection}

\author{Wen-Feng Zhou}
\author{Jun Chen}
\email{jun@pku.edu.cn}
\affiliation{State Key Laboratory for Turbulence and Complex Systems
	\\ Department of Mechanics, College of
	Engineering, Peking University, Beijing 100871, China}

\date{\today}

\begin{abstract}
The corner roll (CR) in the Rayleigh-B\'{e}nard (RB) convection accounts for the behaviors of heat transport and convection flow at the corner. Streamlines of the three-dimensional direct numerical simulations for $10^8<Ra<5\times10^9$ show that CR presents well-defined similarity and multi-layer structure. A stream function for CR is developed by homotopy and the structure ensemble dynamics. The model presents the scaling of Reynolds number of corner roll $Re_{cr}\sim Ra^{1/4}$. Scaling of CR scale $r = 0.77 Ra^{-0.085}$ indicates strong near-wall shearing induced by wind and provides a probability of the `ultimate regime' at high $Ra$. 
\end{abstract}


\maketitle
The Rayleigh-B{\'e}nard (RB) convection is generated in a cell filled with fluid which is cooled from the top plate and heated from the bottom plate. The control parameters of an RB system are the Rayleigh number $Ra = \alpha g\Delta {L_z^3}/(\kappa \nu)$ and the Prandtl number $Pr=\nu/\kappa$ of the fluid, and the aspect ratio $\Gamma=L_x/L_z$, where $g$ is the gravitational acceleration, $\alpha$ the thermal expansion coefficient, $\nu$ the kinematic viscosity, and $\kappa$ the thermal diffusivity, $L_x$ and $L_z$ are the dimensions of the cell in width and height directions, respectively. The response parameters are the Nusselt number, $Nu=q/\left[-\kappa (\partial T/\partial z)\right]$ and the Reynolds number, $Re = UL/\nu$.

The corner roll (CR) is the secondary flow at the corners of an RB cell, induced by the large-scale-circulation (LSC). Usually found in experiments \cite{krishnamurti1981large, qiu1998spatial, niemela2001wind} and numerical simulations \cite{benzi2008numerical, sugiyama2009flow, shi2012boundary}, the CR as a quasi-steady structure at the corner persistently contributes to heat transfer for high $Ra$ number, despite $Pr$ \cite{wagner2012boundary}. Some authors suggested that the local heat transfer coefficient (or the local $Nu$ number) for the CR is larger than that in \textit{the shear region} \cite{kaiser2014local, du2014turbulent}, being explained as the result of strong fluctuation and strong vorticity carried by CR with cool jet from cooling plate impinging on the heated wall \cite{du2016evolution}. In this sense, it is worthwhile to quantify the spatial structure of the CR and its heat transfer performance with a mathematical model. Some evidence indicates that the scale of CR decreases as increasing $Ra$. Thus investigating the behavior of CR also provide a way to explore the physics of the ultimate regime predicted by \citet{kraichnan1962turbulent} in 1962. 

Some researches have attempted to study the CR-type flow based on its properties. For high $Re$ number cases, \citet{batchelor1956steady} first proved that, under the steady Euler limit, for two-dimensional enclosed by vortex sheets, the vorticity away from the sheets is constant. \citet{blythe1995thermally} generalized this work to fluid with body force. For moderate $Re$ number,~\citet{burggraf1966analytical} found that when $Re > 100$, the vortex develops from completely viscous to inviscid rotational core. They also derived that under Euler limit, the temperature in the core is uniform. For small $Re$ number, \citet{moffatt1964viscous} found a series of similarity solutions using separation variable method solving the Stokes flow. He found that for the right angle, any flow sufficiently near the corner must consist of a sequence of eddies decreasing their size and intensity. In mathematical modelling of CR, \citet{batchelor1956steady} and \citet{burggraf1966analytical} proposed the analytical solutions by assuming the circular shape of the CR. However, the conditions for the solutions limit their application to the flow with complex configurations, e.g. the corner region in the RB convection. A solution for small $Re$ number was suggested by \citet{moffatt1964viscous}, but it still cannot be directly applied to an RB cell either in the corner region of relatively high $Re$ number or high $Ra$ number. Therefore, it is necessary to develop a model to describe the CR in turbulent convection. 

The SED theory claims that there are three kinds of basic ansatz in turbulent boundary layer (BL), i.e. power law, defect power law and generalized invariant relation by Lie group analysis \cite{She2017JFM}. Having been examined by canonical wall-bounded turbulence \cite{chen2016bulk, wu2017invariant}, the SED also unifies the temperature profile and the $Ra$-scaling of coefficient of the log-law \cite{she2014prediction}. In applications, the SED has been extended to develop turbulent transition model and RANS model for the flow around foils considering the effects of pressure gradient and finite $Re$ number \cite{xiao2016new}. Being invoked by the geometric similarity of the corner flow and the SED theory, we here developed a self-similarity model for CR based on SED and homotopy analysis.

We performed the 3D DNS for $Ra=10^{8}, 5\times10^{8}, 10^{9}, 5\times10^{9}$ with $Pr=0.7$ of the Boussinesq equations using a second-order staggered finite difference scheme from \citet{verzicco1996finite}. The flow is confined in a narrow rectangular cell with aspect ratio of $L_{x} \text{(width)}: L_{y} \text{(depth)}: L_{z} \text{(height)} = 1:1/6:1$. The periodic boundary condition is employed in the depth ($y$) direction, and the large-scale convection flow is thus held on the $xz$-plane. The resolution of the simulations is up to $1024\times 256 \times 800$ by clustering grid points near the boundaries. The largest grid scale is still smaller than the Kolmogorov and Batchelor scales \cite{shishkina2010boundary, stevens2010radial}, ensuring that both momentum and thermal energy are well resolved. The scaling of $Nu$ of the DNS is $Nu = 3.27 Ra^{0.294}$, consistent with other DNS and experiments \cite{ahlers2009heat, chilla2012new}.

\begin{figure}
	\subfigure[\label{subfig:streamT-1R8}]{\includegraphics[width = 0.45\columnwidth]{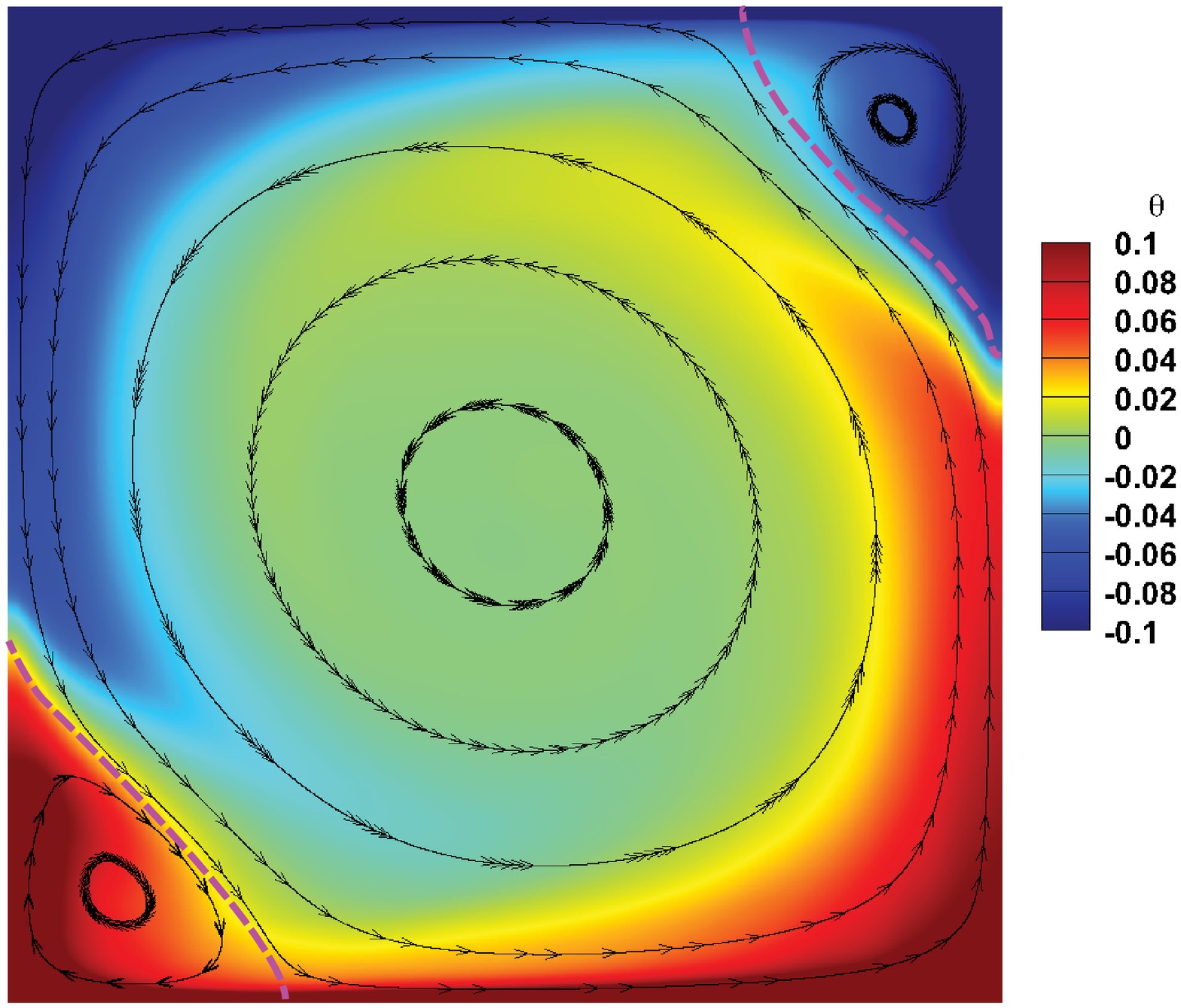}}
	\subfigure[\label{subfig:Re-Ra}]{\includegraphics[width = .54\columnwidth]{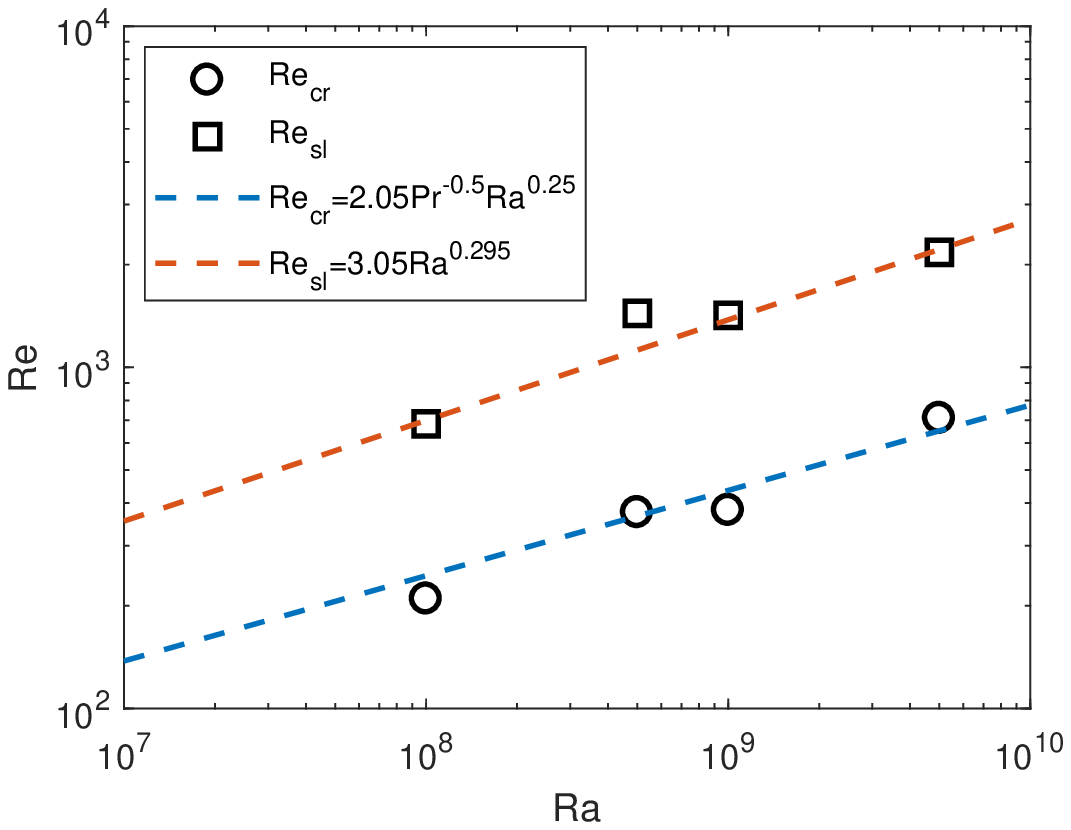}}
	\caption{(a) Time-average field colored by temperature for $Ra=1\times 10^8$. The arrowed-lines represent the streamline. The purple dashed lines mark the slip lines. (b)The $Ra$-scaling of $Re$ of CR, slip line, represented by symbols. The blue line is the CR scaling given by the model. The black line is the empirical fitting of the slip line scaling. The red line is the correlation of CR and the slip line. \label{fig:Ave-Re}}
\end{figure}

Since no \textit{reversal} of LSC occurs through the computational time and the quasi-steady CRs are held at the corners of the cell, the statistical properties of CR are obtained by time-average of flow field, which are shown in Fig. \ref{subfig:streamT-1R8}. The triangular CRs are wedged at the corners along the diagonal direction of the cell. The slip line as the interface between CR and LSC playing a key role in transporting kinetic energy between the large-scale structures. In statistical sense, the energy sustaining rotating of a CR is supported both by shearing of slip line from LSC and the buoyancy from horizontal conducting plate.

It is seen that the $Ra$ number has limited effect on the triangle-shaped CR, the scale of which diminishes with increasing $Ra$. The velocity on the slip line reaches it maximum in the middle with zero velocity at both stagnation ends. Particularly, the maximum velocity in the middle of the slip line implies local intensive momentum and energy transport between CR and LSC, which decreases with increasing $Ra$. To quantify the momentum transfer ability of the slip line, we define the Reynolds number of the slip line as $Re_{sl}=\int\frac{v_{sl}}{\upsilon}\mathrm{d}l$, where $v_{sl}$ is the average velocity on slip line and $l$ the length of slip line, representing a scaling law of $Ra$, $Re_{sl} = 3.15 Ra^{0.295}$; see Fig. \ref{subfig:Re-Ra}.

A model based on the circumferential similarity of the stremlines inside CR is developed. An algorithmic expression of the streamlines is obtained by \textit{homotopy}, giving the transformation expression of two formula continuously deforming from one to another \cite{liao2003beyond}. We here construct a homotopy $H: X\times I \rightarrow Y, I=[0,1]$ for any $f \in X, H(f,0)=f_{2}$ and $H(f,1) = f_{1}$, where $f_{2}(x,z) = 0$ represents a small vicinity of the CR center, and $f_{1}(x,z)=0$ is the function of the CR boundary. When $\xi \in I$ as the similarity variable changes from $0$ to $1$, $f(x,z)=0$ for $H(f,\xi)=0$ deforms continuously from $f_{2}(x,z)=0$ to $f_{1}(x,z)=0$.

The implicit homotopy expression is simplified by linearly combining $f_{1}$ and $f_{2}$, $H=\xi f_{1} + (1-\xi)\Lambda_{0}f_{2}$. If $H=0$, we get the explicit form of $\xi$ as
\begin{eqnarray}
\xi  = {\Lambda_{0}f_{2}}/ ({-f_{1}+\Lambda_{0}f_{2}}) \label{eq:xi}
\end{eqnarray}
where $\Lambda_{0}$ is considered as the constraint strength between the inner and outer layer. 

The least square procedure is used to obtain $f_{1}$, $f_{2}$ and the magnitude $\Lambda_{0}$. The boundary of CR, $f_{1}$, is composed of two walls and the slip line. To keep accuracy, the slip line is expressed as a cubic function $z-(a + bx + cx^{2} + dx^{3})=0$. The walls are respectively written as $x=0$ and $z=0$. Thus the boundary of CR is given as $f_{1}=-xz(z-(a + bx + cx^{2} + dx^{3}))=0$. On the other hand, the inner boundary as a small core at the central region of CR is expressed as $f_{2} = (x-s_{0})^{2} + (z-t_{0})^{2} - \epsilon^2 = 0$, where $(s_{0}, t_{0})$ is the coordinates of the center, and $\epsilon$ the radius of the core. In the case of $Ra = 10^{8}$, they are $a=0.36, b=-1,955, c=10.07, d=-29.58, s_{0}=0.11, t_{0}=0.11, \epsilon=0.0001$. The strength factor $\Lambda_{0}$ in the range of $[0.05, 0.2]$ (e.g. $\Lambda_{0}=0.12$ for $Ra=1\times 10^{8}$) indicates that, for the present cases, the constraint of the rigid wall is always stronger than that of the center.

We consider independent similarity variable $\xi$ (Eq. (\ref{eq:xi})) and the stream function $\psi$ as the dependent similarity variable. The dimensionless Reynolds-averaged Naiver-Stokes (RANS) equations can be written as
\begin{equation}
	u_i\partial_i u_j = -\partial_j p+ Ra^{-1/2} Pr^{1/2} \partial_{ii} u_j - \partial_i\overline{u_j u_i} + \theta \delta_{j3},
\end{equation}
where $u_j, p, \theta$ represent the Reynolds-average quantities. The velocity component is expressed in the stream-function form, using $u_1 = {{\partial \psi }}/{{\partial z}}$, $u_3 =  -{{\partial \psi }}/{{\partial x}}$. Then we have
\begin{equation} \label{Balance}
A=V+\Pi,
\end{equation}
where $A=\partial_3\psi \partial_{ii}\partial_1\psi - \partial_1\psi \partial_{ii}\partial_3\psi$ is advection term,  $V=Ra^{-1/2} Pr^{1/2} \partial_{ii}\partial_{jj} \psi$ is viscous term and $\Pi= \partial_{13}\left(\overline{u'_3u'_3-u'_1u'_1}\right)+\partial_{11}\overline{u'_1u'_3} - \partial_{33}\overline{u'_1u'_3} -\partial_1 \theta$ is fluctuation term. 

As discussed above, the viscosity effect on CR is relatively small and negligible. Thus, the balance is $A \simeq \Pi$. Based on the geometry similarity, we assume that stream function has a similarity solution under the similarity variable $\xi$, i.e. $\psi  = \psi [\xi ] = \psi [x,y;a_0]$, and have
\begin{equation}
\psi_\xi\left(\psi_{\xi\xi} h_1 +\psi_\xi h_2\right) = \Pi,
\end{equation}
where $h_1 = 2[\partial_1\xi\partial_3 \xi \partial_{11}\xi+\left(\partial_3 \xi\right)^2\partial_{13}\xi-\left(\partial_1 \xi\right)^2\partial_{13}\xi - \partial_1\xi\partial_3\xi\partial_{33}\xi]$ and $h_2 = \partial_{111}\xi\partial_3\xi+\partial_{133}\xi\partial_3\xi - \partial_{113}\xi \partial_1\xi -\partial_{333}\xi\partial_1\xi$, representing the coordinate transformation functions. A similarity solution requires $\Pi$ to be a function of $\xi$, i.e. $\Pi=\Pi(\xi)$, and the ratio of $h_2/h_1 = c_2$ is a constant or a function of $\xi$.

The DNS show that the viscosity effect on CR is so weak that negligible. Thus a relationship between the fluctuation terms and the steam-function-differential term $\psi_\xi$ is established as $\Pi/\psi_\xi^2 = c_1$. We have $\psi_{\xi}  \left[\psi_{\xi\xi} + \psi_\xi \left(c_1 + c_2\right) \right] = 0$
with its solution $\psi  = \psi_{0}{\mathrm{exp}(- \xi/\xi_0)} + B$, where $\xi_{0}=1/({c_1} + {c_2})$. Considering the inner boundary condition $\psi(\xi  = 0) = 0$, then the stream-function follows
\begin{equation}
\label{eq:solution}
\psi  = \psi_{0}\left(1 - {e^{ -\xi/\xi_{0} }}\right), 
\end{equation}
as an exact similarity solution. The coefficients $\psi_{0}$ and $\xi_{0}$ can be measured from DNS (e.g. $\psi_{0}=0.028, \xi_{0}=1.9429$ for $Ra=1\times 10^8$). By analyzing the stream function profiles along the similarity coordinate $\xi$ at different central angles, one can see that the stream function collapses well to the DNS, except for the profile of angle of $\pi/6$. Eq. (\ref{eq:solution}) is the solution of the flow under the solid wall condition, thus the profile  near the slip line has a perceptive deviation.
\begin{figure}
	\subfigure[The CR colored by the magnitude of velocity. \label{fig:CR_shape}]{\includegraphics[width=0.4\columnwidth]{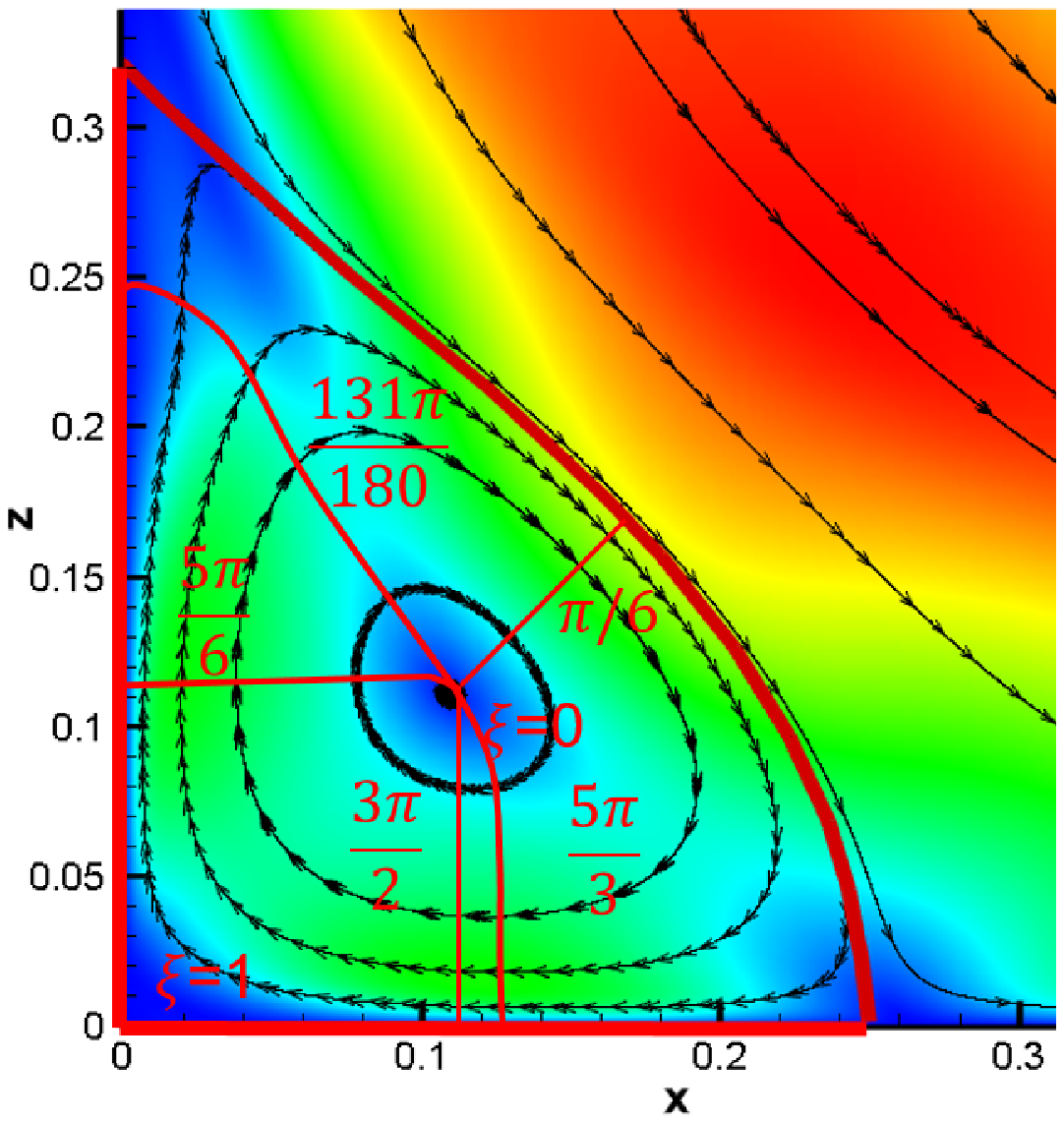}}
	\subfigure[Stream function profile for the angles marked in Fig. \ref{fig:CR_shape}. \label{fig:stream-func}]{\includegraphics[width = 0.59\columnwidth]{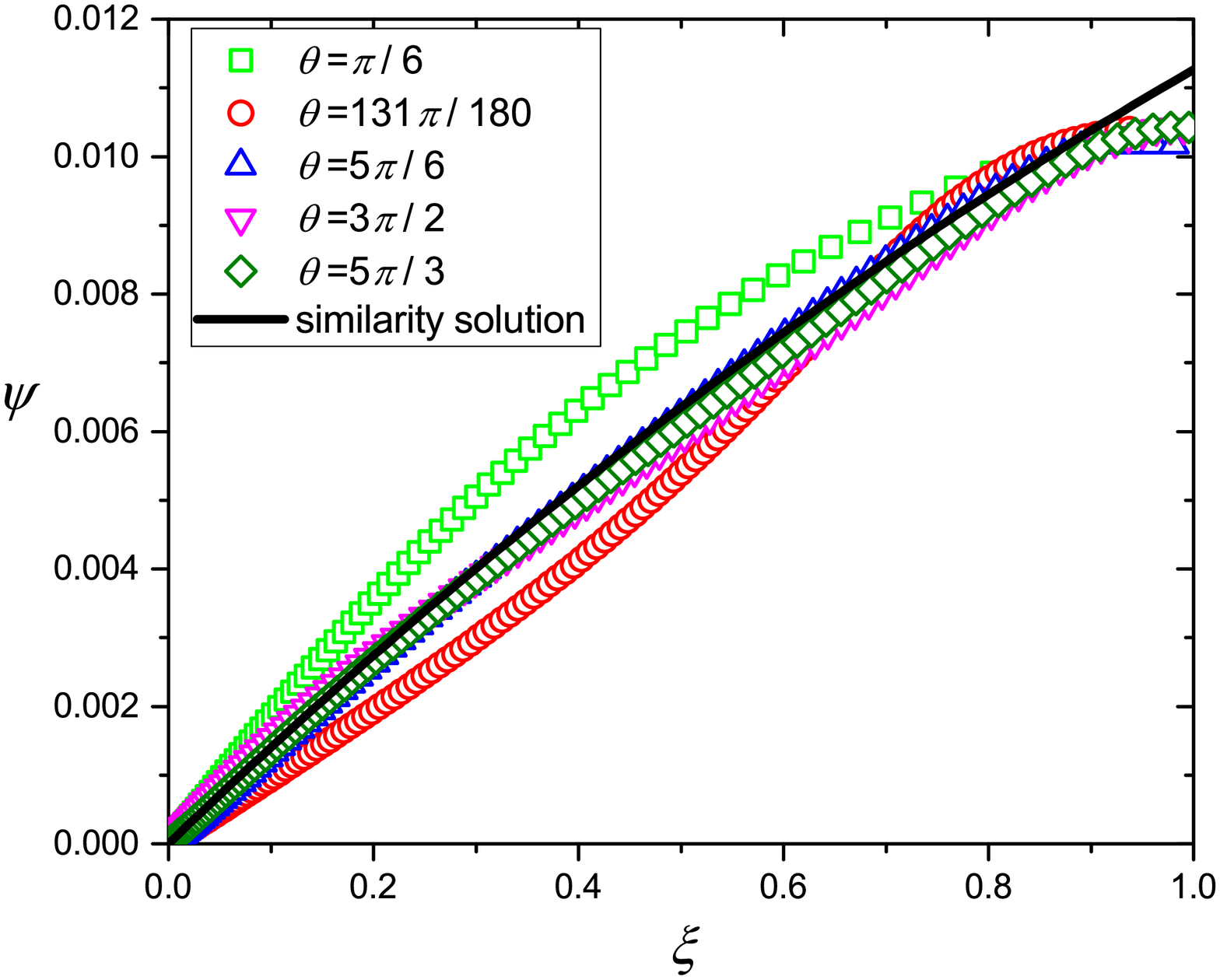}}
	\caption{The CR and its stream function profiles at five angles for the case of $Ra=10^8$.}
\end{figure}

The stream function for the bulk is no longer proper for the near-wall flow --- viscosity should be taken into account. Canonical turbulent BL theory cannot successfully describe the undeveloped flow in CR. Moreover, the fluid near the wall being advected downstream undergoes the varying pressure gradient along with emission of thermal plumes. To obtain a function of boundary layer (BL) in CR, we apply the structure ensemble dynamics (SED) theory \cite{She2017JFM}, by introducing the stress length and the symmetry of the wall taking into account the constraint of solid wall, pressure gradient, and thermal effects \cite{she2014prediction}. 

Space average of the momentum equation in the steamwise direction and integration from $0$ to $z$ gives
\begin{equation}
\label{eq: RANS}
\frac{{d{u^ + }}}{{d{z^ + }}} - \frac{{\overline {u'w'} }}{{u_\tau ^2}} = 1 + \frac{1}{{u_{\tau }^{2}L_{z}}}\int_0^z {({p_r} - {p_l})dz},
\end{equation}
where $u$ and $u'w'$ are steamwise-average variable and $p_r$ and $p_l$ are the right and left side wall pressure, respectively. Velocity and length scale are normalized by friction velocity ${u_\tau } = \sqrt {{{\left. {\nu du/dz} \right|}_{z = 0}}} $ and viscous length ${l_\nu } = \nu /{u_\tau }$. By introducing ${S^ + } = \frac{du^{+}}{dz^{+}}$ ,${W^ + } = - \frac{{\overline {u'w'} }}{u_{\tau}^{2}}$, ${\tau ^ + } = 1 + \frac{1}{{u_{\tau }^{2}L_{z}}}\int_0^z {({p_r} - {p_l})dz} $, we have ${S^ + } + {W^ + } = {\tau ^ + }$.
In the region near the wall, $\tau^{+} \simeq 1$.
Substituting the stress length function $\ell _M^ +  = \sqrt {{W^ + }} /{S^ + }$ as the multilayer similarity function of the BL gives 
\begin{equation}
\label{eq: S}
{S^ + } = \left({{ - 1 + \sqrt {1 + 4\ell _M^{ + 2}{\tau ^ + }} }}\right)/\left({{2\ell _M^{ + 2}}}\right).
\end{equation}
The velocity profile is eventually obtained by integrating from $0$ to $z$, $u = {u_\tau }{u^ + } = {u_\tau }\int_0^z {{S^ + }dz}$.

We apply SED to quantify the pressure gradient effect on BL in CR. The stress length along $x$ direction is $\ell _{M,x}^ +  = \sqrt {{W_x}^ + } /{S_x}^ + $ for the horizontal BL and the stress length in $z$ direction $\ell _{M,z}^ +  = \sqrt {{W_z}^ + } /{S_z}^ + $ for the vertical BL. We apply Taylor expansion to the shear rate $S^+$ and the Reynolds stress $W^+$. The no-slip boundary and continuity condition follow $u \sim z$, $w\sim z^{2}$ for horizontal BL. We have $\ell _{M,x}^ +  = \sqrt {{W_x}^ + } /{S_x}^ +  = {\ell_{0,x}}{z^{3/2}}$ for the BL. Similarly, the stress length of vertical BL is $\ell _{M,z}^ +  = \sqrt {{W_z}^ + } /{S_z}^ +  = {\ell_{0,z}}{(x^+)^{3/2}}$, where $\ell_{0,x}$ and $\ell_{0,z}$ are the coefficient and the function of $x$ or $z$, respectively. The parameter $\ell_{0,x}$ is expressed as follows 
\begin{equation}
\label{eq: rho_x}
{\ell_{0,x}} = \left\{ \begin{array}{l}
a_1(x - x_0),{\rm{  }}0 < x < x_c\\
b_1/(x_{ap} - x),{\rm{  }} x_c < x < x_{ap}
\end{array} \right.
\end{equation}
where $x_0$ ($0.01<x_0<0.0533$ for the present study) is the critical point defining the two ranges; $x_c$ ($0.0747<x_c<0.11$) the position of the core of CR for different $Ra$ numbers. The parameters, $a_1$ ($2.35<a_1<4.5$), $b_1(=0.4Ra^{-0.125})$, $x_{ap}$ ($0.19<x_{ap}<0.26$) are related to $Ra$ number. Similarly, $\ell_{0,z}$ can be expressed as
\begin{equation}
\label{eq: rho_z}
{\ell_{0,z}} = \left\{ \begin{array}{l}
b_3/z, 0< z < z_c\\
a_3(z_0-z), z_c < z < z_{0},
\end{array} \right.
\end{equation}
where $z_0$ ($0.206<z_0<0.2692$ for the present study) is the position of the separation point, $z_c$ ($0.0819<z_c<0.11$) is the position of the core of CR. The parameters, $a_3(=500Ra^{-0.28})$ and  $b_3(=1.3Ra^{-0.2})$ are determined by $Ra$ number. It is noteworthy that the relations between the parameters and $Ra$ number, i.e. $x_0$, $x_c$, $a_1$, $x_{ap}$, $z_0$ and $z_c$ as the functions of $Ra$, cannot be determined with the present simulated cases. Further simulations with wider range of $Ra$ are necessary to obtain the parameters in the stress lengths. Combining the stream function for the bulk and the multi-layer structure for the BL, we have a unified model for CR. $\ell_{0,x}$ and $\ell_{0,z}$ indicate that the eddy scale is inversely proportional to the upstream distance with favorable pressure gradient and decreases linearly with adverse pressure gradient. 

Comparison of the model-reconstructed CR and the DNS, we find that the streamlines and velocity distribution are almost the same except for those very close to the core and the slip line. At the core, low Reynolds effect might cause the elliptical core. The slip line is expressed in Eq. (\ref{eq:xi}), which does not specify the difference between the solid wall and the slip line, though the solid wall has stronger constraint than the slip line. As an assessment of the model, the relative error is examined as $10.93\%$ for $Ra = 1\times10^8$  and $10.45\%$  for $Ra = 10^9$ --- all below $11\%$.
\begin{figure}
	\subfigure[]{\includegraphics[width = 0.49\columnwidth]{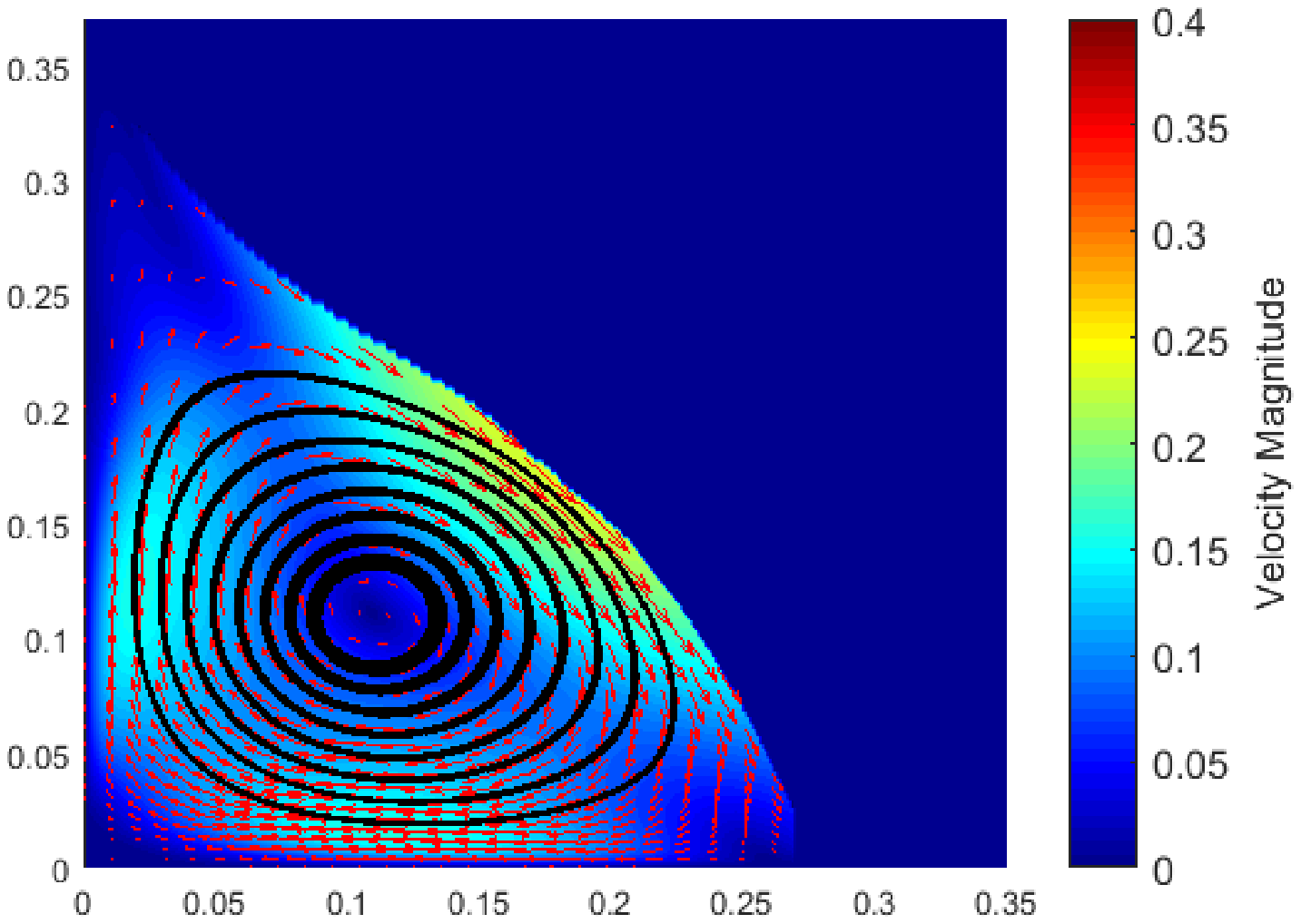}\label{fig:1e8_vDNS_modstre_re}}
    \subfigure[]{\includegraphics[width = .49\columnwidth]{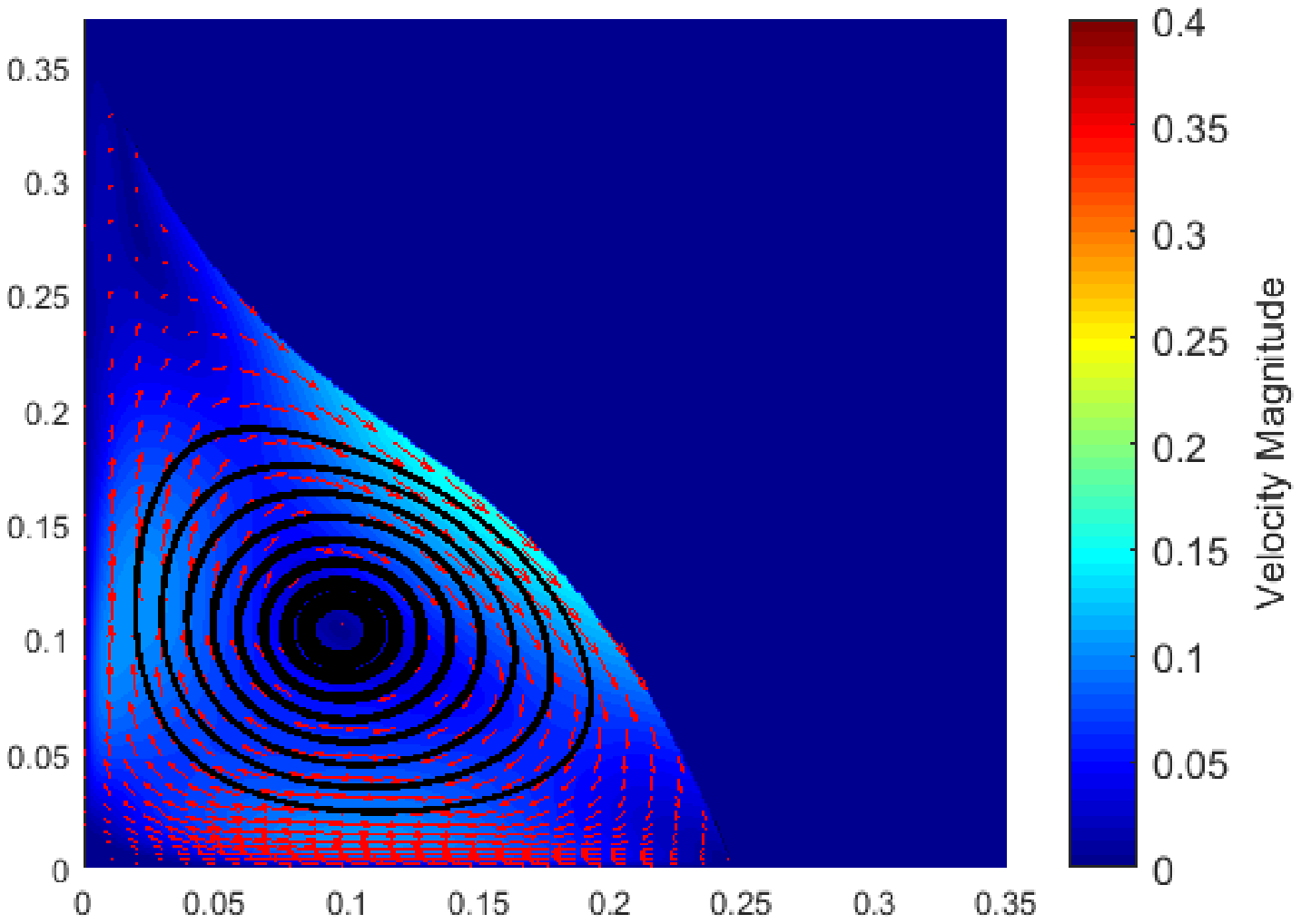}\label{fig:1e9_vDNS_modstre_re}}
	\caption{Comparison of DNS with reconstruction of the similarity solution. Velocity magnitude nephogram and vector arrow represent the DNS and the steam-line is obtained from Eq. (\ref{eq:solution}) (a) $Ra=1\times10^8$; (b) $Ra=10^9$}
\end{figure}

The downwelling flow around the slip line is fully turbulence, and the CR is confined in the corner. As the heated bottom plate pumps energy into the corner region, enforcing the vorticity of CR. Thus, the scale of the CR varying with $Ra$ is detemined by these effects. The CR is characterized both by  $|\psi_{0}/\xi_{0}|$, which is $\psi_{0}/\xi_{0} = 1.45Ra^{ - 1/4}$, and the scale of CR, $r$.

The scale of CR, $r$ is defined as the average distance of the core to the intersection point of walls, which is found to obey the scaling law as $r = 0.77 Ra^{-0.085}$. The negative power exponent indicates that high $Ra$ would suppress the CR and thus elongate the wind shearing region. Recalling the conjecture of realization of `ultimate region' induced by the fully turbulent BL \cite{kraichnan1962turbulent, grossmann2000scaling, grossmann2011multiple}, a relatively wilder shear region would improve turbulence of the BL \cite{tollmien1930entstehung, schlichting1933enstehung}. Therefore, the Reynolds number of CR is given as
\begin{equation}
\label{eq: re_cr}
{Re_{cr,mod }} = \frac{{{u_{cr,mod }}r\cos \theta }}{\nu } \simeq \frac{{\left| \psi_{0}/\xi_{0} \right|}\sqrt{Ra/Pr}}{{C_{1} }}
\end{equation}
Using $r/L_{z}=0.77Ra^{-0.085}$, $|\psi_{0}/\xi_{0}|=1.45Ra^{ - 1/4}$ and $C_{1}=1/\sqrt{2}$ , we have $Re_{cr,mod} = 1.45\sqrt{2}Ra^{1/4}Pr^{1/2}$. It is noted that although the characteristic velocity and scale become smaller at higher $Ra$, the solution of Moffatt for the Stokes flow of the corner cannot be applied either.

The heat transport in the corner roll region is investigated. Being invoked by the hypothesis of mixing zone \cite{castaing1989scaling}, we obtained the correlation between $Re_{cr}$ and $Nu_{cr}$. The assumptions are listed as follows: (a) The characteristic velocity of CR fulfils an anomalous scaling as $u_{cr} \sim (\alpha g \Delta_{cr} L)^{1/2} Ra^{\chi}$; (b) temperature scale of the bulk of CR is proportional to the temperature difference of the top and bottom plate, i.e. $\Delta_{cr}/\Delta \sim const. $; (c) heat flux is determined by the inner convection heat transport, $H \sim u_{cr} \Delta_{cr}$; (d) $w_{h}$ is of the same order of the CR's characteristic velocity, i.e. $u_{cr} \sim w_{h}$; (e) during the emission of plumes, viscous force and buoyancy are balanced as $\alpha g \Delta \sim \nu w_{h}/\lambda^{2}_{cr}$; (f) defining a number of scaling indexes through the relations $Re_{cr}=\frac{u_{cr}r}{\nu} \sim Ra^{\epsilon}, Nu_{cr}^{-1}= \lambda_{cr}/L_{z} \sim \frac{\kappa \Delta/L_{z}}{H} \sim Ra^{-\beta}, \frac{r}{L_{z}} \sim Ra^{\eta} $. Based on relations of (a), (b) and (f), we have 
\begin{equation}
 \label{eq: Nu scaling1}
  \frac{u_{cr}r}{\nu} \sim (\frac{\alpha g \Delta L_{z}^{3}}{\kappa \nu})^{1/2}  (\frac{\kappa}{\nu})^{1/2} Ra^{\chi} \frac{r}{L_{z}}.
\end{equation}
Relations of (c) and (f) follow
\begin{equation}
 \label{eq: Nu scaling2}
 \frac{u_{cr}r}{\nu} \frac{\nu}{\kappa} \frac{L_{z}}{r} \sim Ra^{\beta}.
\end{equation}
Relations of (d), (e) and (f) give
\begin{equation}
  \label{eq: Nu scaling3}
  \frac{u_{cr}r}{\nu} \sim  \frac{\alpha g \Delta L_{z}^{3}}{\kappa \nu} \frac{\lambda_{cr}^{2}}{L_{z}^2} \frac{\kappa}{\nu} \frac{r}{L_{z}}.
\end{equation}
The power exponents in Eq. (\ref{eq: Nu scaling1}) -- (\ref{eq: Nu scaling3}) give
\begin{eqnarray} \label{eq: Nu scaling}
\left\{ \begin{array}{l}
\varepsilon  = 1/2 + \chi  + \eta \\
\beta  = \varepsilon  - \eta \\
\varepsilon  = 1 - 2\beta  + \eta 
\end{array} \right.
.
\end{eqnarray}
With the scaling of CR scale $\eta=-0.085$, we get the rest exponents, which are $\beta = 1/3$, $\epsilon =1/3 - \eta = 0.248$, and $\chi= 1/3 -1/2 = -1/6 \approx -0.167$. The DNS show that the scalings of the characteristic velocity $u_{cr}/U_{f} \sim Ra^{-0.165}$, and that of Reynolds number $Re_{cr}=\frac{u_{cr}r}{\nu} \sim Ra^{0.250}$, close to the scalings from the assumptions. More importantly, the thermal boundary thickness described by the approximation is $\lambda_{cr}/L \sim Ra^{-1/3}$, in agreement with the DNS, $\lambda_{cr,dns}/L_{z} \sim Ra^{-0.331}$. It is noteworthy that the assumption of the constant CR temperature has only been examined with the unit aspect ratio. The validity of the assumption at different aspect ratios needs further investigation.

The present results can be extended to the large scale circulation. While we established the stream function for corner roll, it is likely that the stream-function for the boundary for the triangular shape of the corner roll but also applicable for the hexagon boundary of the bulk flow; see Fig. \ref{subfig:streamT-1R8}. Another important aspect is the multilayer structure of the vertical and horizontal profiles near the walls. We anticipate that, depending on the boundary type, the boundary layer of the large-scale circulation may be described by the similar multilayer function, in particular for the solid wall boundary, and for the slip line.

Another intriguing question relates to the assumptions specified for the CR, in relation to the $Re$ number. The positive scaling of $Re_{cr}$ ($\epsilon = 0.248$) indicates that the corner roll is enforced by increasing the $Ra$ number, though the size of CR decreases with the $Ra$ number with a negative exponent $\eta = -0.085$. The enhancement of convection leads intensive heat transport at higher $Ra$ number, i.e. $Nu\sim Ra^{0.331}$, where the power exponent even higher than in the wind shearing region ($Nu\sim Ra^{0.295}$).

From the hypothesis building perspective, our work also motivates investigating separation scenarios leading to enhanced heat transport. In the future, we also plan to explore parameterization of the streamline functions of CR and bulk flow. Finally, it may prove fruitful for more detailed simulations to study the corner roll under various control parameters, such as the $Pr$R number, $\Gamma$.

We thank Zhen-Su She for helpful comments and suggestions. This work is supported by National Nature Science (China) Fund 11452002, 11521091, and 11372362, and by MOST (China) 973 project 2009CB724100.

\bibliographystyle{unsrtnat}

\bibliography{CR-instructions}

\end{document}